# The principle of creating quasiperiodic surfaces under the action of vibrating dielectric matrix

D.Koroliouk*, M.Zozyuk and Yu.I.Yakymenko

Faculty of Microelectronics
National Technical University of Ukraine
"Igor Sikorsky" Kyiv Polytechnic Institute
Kyiv, Ukraine

*Abstract*. **A method for creating metasurfaces using a standing wave, formed in dielectric, is proposed. Such metasurfaces are formed from metal suspensions, deposited on dielectric plate, placed in a metal frame-screen. A series of parameters for creating the standing waves is discussed. Here the method for creating Hladni figures using acoustic standing waves is transferred to electromagnetic ones. The mathematical description of this model is proposed, as well as the research methods.**

Metasurfaces are used in filters, resonators, converters of new generation and are studied with great intensity. A method for creating meta-surfaces using vibrating dielectric plate with specially applied surface material is presented. A standing wave of a certain frequency is formed in the dielectric matrix, which causes the material to move on the surface of the dielectric. Such a mechanism has been successfully demonstrated by Ernst Hladni, and the figures created using this method are called "Hladni figures" [1]. This paper presents the principle for creating meta-surfaces based on electromagnetic standing waves. The main parameters of physical model of the proposed process are discussed, used both in physical experiments and in mathematical modeling.

The surface reliefs created by this method can be used in microwave and nanowave technology as filters, resonators, and other wave devices. In this case, it is possible to vary the process parameters in order to get different scales of the surface morphology.

## I. Introduction.

The use of acoustic exciting action, aimed to create original structures, is based on creating of standing wave that form on a base plate, onto which elastic material is applied or strewed. Such a material, under mechanic action of the standing wave, will redistribute its mass, according the position of nodes and antinodes of the standing wave along the plate.

This work aims to describe the methods of selecting materials for creating electromagnetic standing waves when creating surface reliefs similar to Hladni figures (Fig. 1) using non-acoustic method, namely using electromagnetic standing wave in dielectric matrix.

The use of electromagnetic waves in a dielectric matrix causes different polarization mechanisms. For each dielectric, the polarization frequency is different.

It is important to understand that the propagation of electromagnetic waves in a dielectric causes various polarization mechanisms, and this is the most important aspect when choosing a dielectric.

An important task is optimal choice of the relief-forming material that will be applied on the dielectric. A number of such factors must be taken into account: susceptibility to electron plasma, wettability, melting point, isomorphism and others.

Our goal is to study various criteria on the selection of materials with different properties, as well as methods for various parameters evaluation, in order to create metasurfaces using the method of vibrating dielectric matrix.

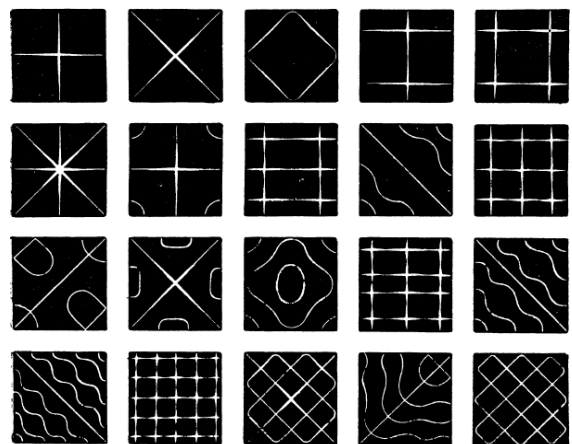

Fig. 1 Hladni figures at different frequencies

In this case, various factors of the physical model should be taken into account, such as diffusion processes under amorphous state of materials, isomorphism of materials, thickness of layers of materials, methods of applying materials at different scales, the degree of purity of the medium, the conditions that will make it possible to study the surface structure ("drawings") other.

## II. Selection of the desired dielectric medium

The problem of the optimal choice of dielectric should be solved basing on the frequency of standing wave. And this,

*Corresponding author *dimitri.koroliouk@ukr.net*

in turn, substantially determines the morphology of metasurface.

The optimal choice of dielectric should be based on the frequency of standing wave. This determines the morphology of metasurface, which is obtained.

The basic formula for calculating the dielectric matrix is

$$\varepsilon(\omega) = 1 - \frac{4\pi Ne^2}{m\omega^2}, \quad (1)$$

where $N$ is the number of electrons in all atoms per unit volume of the substance, $e$ is the electron charge, $m$ is the electron mass, $\varepsilon' + i\varepsilon''$ is the dielectric constant of the dielectric plate.

Formula (1) allows us to calculate the dielectric constant basing on the field frequency, as well as the loss tangent:

$$Tg\delta = \frac{|\varepsilon''|}{\varepsilon'} = \frac{\sigma}{\omega\varepsilon_0\varepsilon'}, \quad (2)$$

where $\varepsilon''$ is the imaginary component of the dielectric constant, $\varepsilon'$ is the real component, $\sigma$ is the conductivity [2].

Of course, the losses can be compensated by the field intensity, but this does not mean that the dielectric will contain a standing wave with the necessary functional characteristics, since it is necessary that the field intensity is sufficient to move the material mass along the surface of the dielectric.

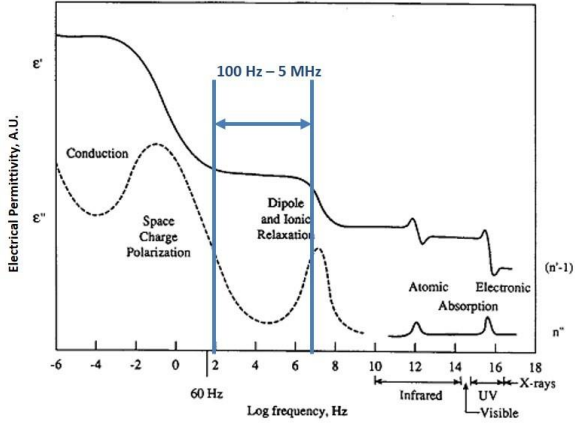

Fig. 2. The dielectric constant components

Figure 2 illustrates typical behavior of the real and imaginary parts of the dielectric constant in a wide frequency range [3].

The regions of sharp changes in these components correspond to absorption lines, which can have different nature: dipole or ion relaxation, atomic and electron resonances at high frequencies [2].

As an example, there are materials based on multi-walled carbon nanotubes with various diameters and morphologies [4]. To study nanotubes, a polyethylene matrix is used, in which nanotubes are embedded. In [5], the real and imaginary components of permeability were shown in the range of 2–4 at frequencies from 0.2 to 2.8 THz.

Of interest is the possible use of dielectric matrices in semiconductors. In silicon crystals, electron polarization is present due to the existence of valence electrons. In [6], the results of studies of the electron and ion polarization of various silicate-containing materials are presented, which have promising possibilities for the experimental study of dielectric matrices.

There is also a great potential for liquids and gases in which the dielectric constant stays in a certain limit (1-1.5). For example, Argon in liquid state (T <87.3) has the permeability within 1.5, while with increasing temperature it will drop down to 1.

The behavior of joint characteristics (parameters) of materials is essential. To study them, the dependences of permeability on the frequency, temperature, layer thickness, number of defects and purity will be investigated. It is assumed that covariance functions of equilibrium (stationary) random processes are used to statistically identify the dependence of some parameters on others.

First of all, we are interested in electron and ion polarization depending on frequency and temperature. Next, dielectric losses, which may be associated with an increase in temperature and frequency, should be studied. In addition, it is necessary to study the compatibility of the dielectric material with important experimental parameters, such as alloying, wettability, and others.

It is important to analyze the electron and ion polarization as a function of frequency and temperature. Then, dielectric losses, which may be associated with an increase in temperature and frequency, should be studied. It is also necessary to study the compatibility of dielectric material with important experimental parameters such as alloying, wettability, and others.

### III. Surface layer material selection

A material to be selected should be susceptible to the external electromagnetic field and at the same time smoothly and easily moving along the surface, to be able to repeat the shape of a standing wave (nodes and antinodes). Here, such parameters as the volume of electron plasma (the number of free electrons), the melting point, and the magnetic and electric susceptibility are important. Depending on which component of the wave will be used to move the material, the selection of the appropriate surface material will be implemented.

For the electrical component of the standing wave, a material with a high electrical susceptibility must be used so that the shielding length is sufficient to capture as many electrons in the surface layer as possible, ideally to capture the entire volume of the material.

The material is selected according to the following criteria: calculation of the screening length, "skin layer", melting temperature, degree of wettability, fusion, temperature and frequency dependence of all components.

So, if the electrical component is used, then the shielding length should be calculated:

$$l_e = \sqrt{\varepsilon\varepsilon_0 kT/e^2 n_0}, \quad (3)$$

where $n_0$ is concentration of electrons in the initial state, $T$ is temperature, $k$ is Boltzmann constant.

Formula (8) determines the depth of penetration of the electric field into the material.

The nodes and antinodes form a force field configuration that controls the charges inside the surface layer and ultimately ensures the surface mass redistribution, together with the charge, of the material.

To describe this process, it is proposed to use the well-known formulas [10] for the amplitudes of the electric and magnetic components at the interface:

$$E_m(z) = E_1^+ |1 + \dot{\rho} e^{i2kz}|, \quad (4)$$

$$H_m(z) = E_1^+ |1 - \dot{\rho} e^{i2kz}| / Z_w, \quad (5)$$

where $k$ is the wave number, $z$ is the coordinate in space, $E_1$ is the amplitude of the electric component at the origin, $\dot{\rho}$ is the reflection coefficient, $Z_w$ is the wave resistance of the environment.

Using these formulas, the behavior of the waves in dielectric environment can be simulated and the location of a standing wave in the dielectric matrix can be determined ("drawn").

As promising materials there are alloys based on magnesium (Mg), gallium (Ga), cadmium (Cd), tin (Sn), bismuth (Vi) and others [8]. It is important to keep in mind alloys based on metals at the top of electrochemical series of activity and having a low melting point [8]. The most suitable metals are magnesium, aluminum, zinc, but difference from metals with strong chemical activity, such as sodium, potassium and other alkali metals.

By analogy with calculation of the material for electrical component, the calculation of material for the magnetic component consists of the following steps: determination of the screening length of magnetic field, magnetic susceptibility of the metal, degree of wettability, fusion, temperature and frequency dependence of the components of material. There are materials based on ferromagnetic fluids inside some neutral environment such as polyacrylic acid, soya lecithin, water, sodium polyacrylate, glycerol and others.

There is a number of other materials of interest due to their low melting point and at the same time with good magnetic properties: paramagnets - cesium, aluminum, magnesium and others (more active), mixtures of ferromagnets and ferrites, as well as alloys based on them.

In [9], it was demonstrated that with increasing the temperature, the magnetic susceptibility increases and the same materials have a low melting point.

The metals can be mixed with liquid substances to achieve special electromagnetic properties: metal powder mixed with glycerin, isoparaffins, transformer oil, metals with a low melting point (gallium, tin, bismuth, indium, as well as their alloys).

It is important to consider other parameters, for example, fusion temperature of materials, as well as via wettability. The calculation of the components becomes more complicated as various experimental factors grow. A study is planned on of the permeability dependence by variotion of temperature, material purity, number of defects, layer thickness and other electrophysical parameters. A statistical analysis of experimental configuration dependence on the chosen parameters will be also investigated.

A large amount of researches consider boundary conditions for the study of model. Such restrictions permit to focus the model parameters in the domain of interest.

### IV. The dynamics of the constituent mechanisms of the experiment

The layer thickness is determined by the frequency of standing wave. The main criterion is that the entire volume of surface layer is used by the field, and therefore, the effective movement of the material is provided.

Depending on the wave frequency, on layer thickness, roughly on the scaling of the experiment, different methods for applying the layers will be implemented. It is necessary to provide a number of preparatory steps, such as cutting a dielectric to create boundary conditions for a waveguide-dielectric, applying a transition layer at the interface between dielectric and surface layer, in order to avoid fusion of boundary materials, which should be avoided in any way (Fig. 3). Therefore, the configuration of the dielectric matrix substantially depends on the parameters of the physical model [10].

To take into account the priority of the parameters, it is necessary to determine which processes are more valuable for the successful conduct of the experiment, and which in the end may not affect the morphology of the surface relief at all.

Firstly, it is necessary to take into account in this context that these are processes that accompany an experiment at an active stage. In any case, diffusion processes will participate, since the operating temperatures can be higher than some limit temperature, and, accordingly, the melting effects will grow disproportionately fast.

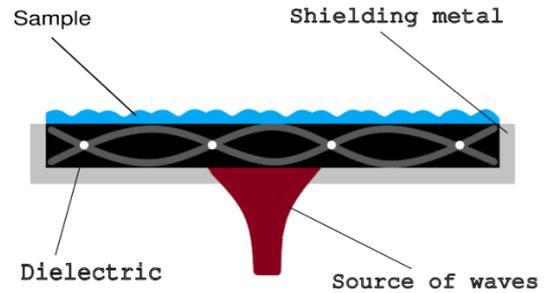

Fig. 3. Schematic installation

The above setups are associated with the former state of a standing wave and a fixed morphology of the metasurface. A change in the characteristics of the dielectric matrix, as well as the parameters of the emitted electromagnetic wave, implies a redistribution of surface masses and a significant change in the morphology of the surface of the metasern. It is important to understand the practical significance of such "smooth regulation". The dynamics of changes in the morphology of the metasurface is described as a stationary diffusion process, which is a solution to the difference stochastic equation:

$$\Delta\alpha_{t+1} = -V\alpha_t + \sigma\Delta W_{t+1}, \; t \geq 0, \quad (6)$$

where $0 < V < 1$; $\alpha_t$, $t \geq 0$ is the "proportional" thickness of the metal layer at a fixed point of the dielectric matrix, $\Delta W_{t+1}$, $t \geq 0$, is the standard Wiener process (Brownian motion) with mathematical expectation 0 and standard deviation 1.

The drift and diffusion parameters $V$, $\sigma$ characterize the dynamics of the process $\alpha_t$, $t \geq 0$ [11]. An important property of the diffusion model (6) is the conditions of the statistical equilibrium of the system expressed in terms of the regression functions [12], which makes it possible to constructively pose the problem of establishing relations between the parameters $V$, $\sigma$ and the parameters $\varepsilon$, $\mu$, $\lambda$, as well as with other characteristics of the physical model for creating metasurfaces using a standing wave.

The dynamic diffusion model also makes it possible to numerically simulate [13] the solutions of the difference stochastic equation (6) with given parameters for a preliminary analysis of the dynamics of the metal layer thickness with a change in the basic parameters of the physical model.

A more complex approach to studying the dynamics of changes in the surface morphology of the metasurface with changing parameters of the physical model $\varepsilon$, $\mu$, $\lambda$ and some matrix configuration parameters can be mathematically modeled as a dynamic diffusion model (7), generalized to the case of discrete diffusion in a random or deterministic medium [14]. That is, the coefficients of the difference stochastic equation depend on the parameters of the external environment $x(t)$ and the dynamic equation becomes:

$$\Delta\alpha_\delta(t) = -\delta^2 V(x(t))\alpha_\delta(t) + \delta\sigma\Delta W(t), \quad (7)$$

where the scaled increment is defined as $\Delta\alpha_\delta(t) \coloneqq \alpha_\delta(t + \delta^2) - \alpha_\delta(t)$, $\delta$ is a small scaling parameter, $\delta \to 0$.

Another model for changing the surface morphology of the metasurface is the consideration of discrete diffusion in a balanced medium [14], which corresponds to a stable state. That is, the leading parameter is $V(x)$, $x \in E$, which can take both positive and negative values, provided that the random environment is balanced:

$$\int_E \rho(dx)V(x) = 0, \quad (8)$$

where $\rho(dx)$ is the invariant measure of the external influence.

Then the difference stochastic equation has the form

$$\Delta\alpha(t) = -\hat{V}\alpha(t) + \hat{\sigma}\Delta W(t), \quad (9)$$

where the coefficients of the equation are the following:

$$V_0 = V_0^2 - V^2, \; \sigma_0^2 = \sigma^2 + (2V_0^2 - V^2)s^2,$$
$$V_0^2 \coloneqq \int_E \rho(dx)V(x)\mathbb{R}_0 V(x), \; V^2 \coloneqq \int_E \rho(dx)V^2(x),$$
$$\sigma^2 \coloneqq \int_E \rho(dx)\sigma^2(x).$$

where the operator $\mathbb{R}_0$ is the generalized inverse operator of $\mathbb{Q} = \mathbb{P} - \mathbb{I}$, that is $\mathbb{Q}\mathbb{R}_0 = \mathbb{R}_0\mathbb{Q} = \Pi - \mathbb{I}$,
where $\mathbb{P}$ is the transient state operator of the external influence, $\mathbb{I}$ is the identity operator and $\Pi$ is the projector onto the subspace of zeros of the operator $\mathbb{Q}$, defined by the equality

$$\Pi\varphi(x) = \int_E \rho(dx)\varphi(x).$$

To take into account transient processes and relaxation, one can also use the equation with asymptotically infinitesimal diffusion [15], namely

$$\Delta\alpha_\delta(t) = -\delta^3 V(x(t))\alpha_\delta(t) + \sqrt{\delta}\sigma\Delta W(t), \quad (10)$$

where the scaled increment is defined as $\Delta\alpha_\delta(t) \coloneqq \alpha_\delta(t + \delta^3) - \alpha_\delta(t)$, $\delta$ is a small scaling parameter, $\delta \to 0$.

## V. Conclusion

The basic components of an experiment to create quasiperiodic matrices based on a vibrating dielectric plate are analyzed.

Directions for research and selection of materials for their use in the experiment are indicated. When finding the optimal dielectric, one must pay attention to the polarization components, since different polarization mechanisms take place at different frequencies.

When selecting the substrate material, it is necessary to determine which component will be used in the transformation of the surface layer. So, when using electrical component, the amount of electron plasma in material must be taken into account, and when using the magnetic component, temperature dependence of dielectric constant on frequency must be taken into account.

In the context of selection of materials, a preliminary study of all components for the compatibility and suitability of their use in the design ranges of parameters is also necessary.

Also, mathematical methods are proposed for studying diffusion processes to understand the consequences of isomorphism of materials, and they will also make it possible to study the dynamics of material movements on the surface of a plate.

**Funding**

This research received no external funding.

**Conflict of interest**

The authors declare that they have no conflict of interest.

**References**

[1] Faraday, M. On a Peculiar Class of Acoustical Figures; and on Certain Forms Assumed by Groups of Particles upon Vibrating Elastic Surfaces. Philosophical Transactions of the Royal Society of London, 1831, v. 121, 299−340.

[2] Michel Houssa. High-k Gate Dielectrics, — CRC Press, 2004. — 601 p.

[3] L.D. Landau, E.M. Lifshitz, "Course of Theoretical Physics. Electrodynamics of Continuous Media. 2nd edition," Butterworth-Heinemann, 1979. vol. 8, pp. 460,

[4] V.L Kuznetsov, I.A. Simonova, A.I. Stadnichenko, A.V. Ishchenko, "Oxidation behavior of multiwall carbon nanotubes with different diameters and morphology," Appl. Surf. Sci., 2012, vol. 12, pp. 258–298.

[5] Tae-In Jeon, Joo-Hiuk Son, Kay Hyeok An, Young Hee Lee, and Young Seak Lee, Terahertz absorption and dispersion of fluorine-doped single-walled carbon nanotube. Journal of Applied Physics., 1998, v. 10, pp. 3-4.

[6] A.C. Lasaga, R.T. Cygan, "Electronic and ionic polarizabilities of silicate minerals," American Mineralogist, 1982., vol. 67, pp. 328–334.

[7] D.R. Lide CRC Handbook of Chemistry and Physics (87th ed.). Boca Raton, FL: CRC Press. 2006.

[8] Y.V. Tolstobrov, N.A Manakov, M.V. Pletneva, "The phenomenon of thermal magnetization of highly anisotropic single crystals," Tech. Phys. Lett., April 2006, vol. 32, pp. 332 – 327.

[9] M.V. Strikha, A.I. Kurchak, A.N. Morozovska, "Influence of Domain Structure in Ferroelectric Substrate on Graphene Conductance," (Authors' Review), Ukr. J. Phys, 2018, vol.63, pp. 49–69.

[10] M.Zozyuk, D.Koroliouk, V.Moskaliuk, A.Yurikov and Yu.Yakymenko. Creation of quasiperiodic surfaces under the action of vibrating dielectric matrices, ELNANO-2020, pp. 224 - 229, DOI: 10.1109/ELNANO50318.2020

[11] D. Koroliouk, "Two component binary statistical experiments with persistent linear regression", Theory of Probability and Mathematical Statistics, 2015, v.90, pp.103-114.

[12] Koroliouk D., Stationary statistical experiments and the optimal estimator for a predictable component /. // Journal of Mathematical Sciences, 2016, 214(2), pp.220-228.

[13] Koroliouk, D.V., Koroliuk, V.S., Rosato, N., Equilibrium Processes in Biomedical Data Analysis: The Wright–Fisher Model. Cybernetics and Systems Analysis, 2014, 50(6), pp.890-897.

[14] Koroliouk, D., Statistical experiments in a balanced Markov random environment. Cybernetics and Systems Analysis, 2015, vol. 51, No. 5, pp.766-771.

[15] Koroliouk D., The problem of discrete Markov diffusion leaving an interval. Cybernetics and Systems Analysis, 2016, vol. 52, No. 4, pp.571-576.